\documentstyle{mn}

\input epsf

\title[Modelling The Two Plane Lens B2114+022]
{Modelling The First Probable Two Plane Lens System B2114+022: 
Reproducing Two Compact Radio Cores A,~D}

\author[K. Chae et al.]
{Kyu-Hyun Chae,$^1$ Shude Mao,$^1$ Pedro Augusto$^2$
 \smallskip  \\  
$^1$University of Manchester, Jodrell Bank Observatory, 
Macclesfield, Cheshire SK11 9DL, UK \\
$^2$Universidade da Madeira, Dep.\ Matem\'atica, Caminho da Penteada, 
9050 Funchal, Portugal}

\date{Accepted ........
      Received .......;
      in original form .......}

\pubyear{2001}

\begin{document}
\maketitle
\begin{abstract}
We test possible lensing scenarios of the JVAS system B2114+022, in which two
galaxies at different redshifts (``G1'' at $z_1 = 0.3157$ and ``G2'' at 
$z_2 = 0.5883$) are found within 2 arcseconds of quadruple radio sources. 
For our investigation, we use possible lensing constraints derived from
a wealth of data on the radio sources obtained with VLA, MERLIN, VLBA and
EVN as well as HST imaging data on the two galaxies, which are presented
in Augusto et al. In the present study, we focus on 
reproducing the widest separated, observationally similar radio components
A and D as lensed images. We first treat G2 
(which is the more distant one from the geometric centre) as a shear term,
and then consider two plane lensing explicitly including G2's potential at
the $z_2$ plane as the first case of two plane lens modelling. Our modelling
results not only support the hypothesis that the system includes
gravitationally lensed images of a higher redshift extragalactic object,
but they also show that the explicit inclusion of G2's potential at the
second lens plane is necessary in order to fit the data with astrophysically
plausible galaxy parameters. Finally, we illustrate a natural consequence of
a two-plane lens system, namely the prediction of distortion as well as 
shift and stretching of G2's isophotes by G1's potential, which can in 
principle be measured by subtracting out G1's light distribution in a high 
S/N and good angular resolution image, especially a multi-colour one.
\end{abstract}


\section{INTRODUCTION}
\setcounter{figure}{0}
\setcounter{table}{0}

At the time of writing (December 2000),
there are $\sim 60$ confirmed or candidate multiply-imaged
extragalactic sources. All these systems except possibly for B2114+022 appear
to have lensing object(s) at a single redshift, although various astrophysical
perturbations may exist at different redshifts (Keeton, Kochanek \& Seljak
1997). The lens candidate system B2114+022 was found by Augusto et al.\ (2000;
hereafter A00)
as the sixth strong lens candidate in the Jodrell-VLA Astrometric Survey
(JVAS) which along with the follow-up Cosmic Lens All Sky Survey (CLASS)
discovered 19 new lenses or lens candidates (e.g.\ Browne 2000). The field
of the JVAS system B2114+022 includes two early type galaxies at 
different redshifts ($z = 0.3157$ and 0.5883) separated by
$\approx 1\farcs3$, and four radio sources two of which are separated by
$\approx 2\farcs6$ enclosing the two galaxies within the diameter. These two
widest separated components (A and D) have similar radio imaging and spectral
properties and are most likely to be lensed images, while the other two 
components are open to several alternative astrophysical origins possibly
including lensing, based upon presently available data (see A00). 
In this paper, we consider a realistic two plane lens 
model for B2114+022~A,~D.

The basic equations of multiple plane lensing are well known, 
and can be derived elegantly from Fermat's principle
(Blandford \& Narayan 1986; Kovner 1987). In general, the properties of
multiple plane lensing are much more complicated than those
of single plane lensing, although some single-plane theories remain
valid in the case of multiple plane lenses (Seitz \& Schneider 1992). 
Due in part to this complexity, only relatively simple two-plane deflectors
have been studied so far. Erdl \& Schneider (1993) gave a complete
classification of the critical curves and caustics for two point lenses 
distributed in different planes, while Kochanek \& Apostolakis (1988) 
investigated the lensing properties of two spherical deflectors at different 
redshifts. The properties of two plane lensing
by elliptical deflectors (e.g.\ caustic properties) 
are essentially unknown. This paper is in part an investigation
of the properties of two plane lensing by elliptical deflectors 
as example models of B2114+022.

Kochanek \& Apostolakis (1988), using the above model,
predicts that 1--10 per cent of gravitational 
lenses should be two plane lenses. This theoretical
prediction is broadly consistent with the statistics of the well-defined,
complete CLASS survey, i.e. one two-plane lens candidate out of $19$.
The fact that no additional two plane lens was discovered in the rest of 
$\sim 40$ more heterogeneous lenses hints that other (less well-defined)
surveys may be biased against the discovery of multiple plane lenses. 
Nevertheless, as the number of observed lenses increases rapidly, 
we expect that more two plane 
lenses will be discovered in the future allowing us to use them for 
astrophysical applications (e.g.\ galactic structures and evolution,
constraining cosmological parameters). 

This paper is organized as follows. In section~2, we briefly review the theory
of two plane lensing and summarize the equations. In section~3, we investigate
possible lens models of B2114+022~A,~D, with particular emphasis on the
differences between the properties of a single-plane lens model and a two-plane
lens model. In section~4, we discuss our results and point out directions for 
future work.

\section{Review of Two Plane Lensing}

In this section, we review the theory of two plane lensing. For further
review, the reader is referred to the monograph by Schneider, Ehlers
\& Falco (1992).  

Let $\vec{\xi}_1$, $\vec{\xi}_2$ and $\vec{\eta}$ be the physical vectors
from a fiducial perpendicular line on the foreground lens, background lens and
source planes, respectively (Figure~1). The fiducial line, called the optical 
axis, is defined in this paper as the line passing through the mass centre of
the foreground deflector (Figure~1). 
\begin{figure*}
\begin{center}
\setlength{\unitlength}{1cm}
\begin{picture}(17,17)(0,0)
\put(2.,-0.5){\includegraphics{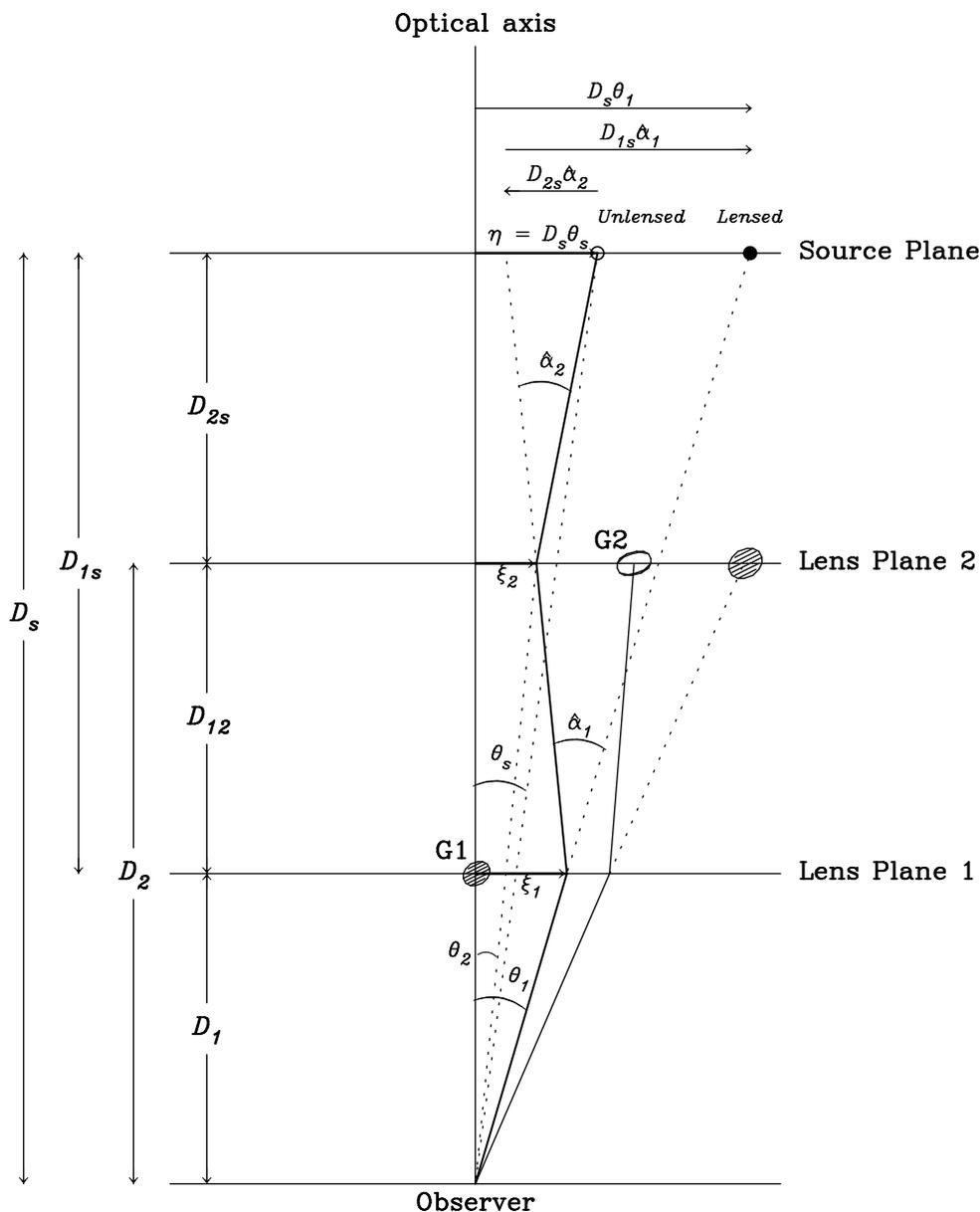}}
\end{picture}
\caption{
Source and image positions and ray-paths for two plane lensing. 
The physical quantities used in section~2 are indicated. The shift of
the background lens's position due to the foreground lens is illustrated.
Although not shown in the diagram, isophotes of the background galaxy
are stretched and distorted resulting in apparent change in the galaxy's
position angle, ellipticity and isophotal shapes.
}
\label{}
\end{center}
\end{figure*}
Let $\hat{\alpha}_i(\vec{\xi}_i)$ 
($i =1,2$) be the deflection angles due to the foreground and background
deflectors respectively. The impact vector ($\vec{\xi}_1$) on the foreground
lens plane is related to the physical source vector ($\vec{\eta}$) and the
impact vector ($\vec{\xi}_2$) on the background lens plane by the following
two plane and single plane lens equations:
\begin{equation}
\vec{\eta} = \frac{D_s}{D_1} \vec{\xi}_1 - D_{1s} 
\hat{\alpha}_1(\vec{\xi}_1)
   - D_{2s} \hat{\alpha}_2(\vec{\xi}_2)
\end{equation}
and
\begin{equation}
\vec{\xi}_2 = \frac{D_2}{D_1} \vec{\xi}_1 - D_{12} \hat{\alpha}_1(\vec{\xi}_1).
\end{equation}
In the above all the distances are angular diameter distances. As for the
single plane lens case, it is convenient to re-scale the physical vectors
by length units in proportion to the angular diameter distances to the three 
planes from the observer, i.e., by $\xi_0$ (an arbitrary length), 
$\xi_0 D_2/D_1$ and  $\xi_0 D_s/D_1$; in other words, we define 
$\vec{x}_1 \equiv \vec{\xi}_1/\xi_0$,
$\vec{x}_2 \equiv \vec{\xi}_2/(\xi_0 D_2/D_1)$ and
$\vec{x}_s \equiv \vec{\eta}/(\xi_0 D_s/D_1)$ (e.g.\ for $\xi_0 = D_1$,
they become angle vectors). Using these dimensionless scaled quantities, 
the lens equations (1) and (2) become
\begin{equation}
\vec{x}_s = \vec{x}_1 - \vec{\alpha}_1(\vec{x}_1) - \vec{\alpha}_2(\vec{x}_2)
\end{equation}
and
\begin{equation}
\vec{x}_2 = \vec{x}_1 - \beta_{12} \vec{\alpha}_1(\vec{x}_1),
\end{equation}
where 
\begin{equation}
\vec{\alpha}_i(\vec{x}_i) = \nabla \psi_i (\vec{x}_i) \hspace{1.0cm} (i=1,2)
\end{equation}
are scaled dimensionless deflection angles, and
\begin{equation}
\beta_{12} = \frac{D_{12} D_s}{D_2 D_{1s}}.
\end{equation}
In equation (5) the dimensionless potentials are given by
\begin{equation}
\psi_i(\vec{x}_i) = \frac{1}{\pi} \int d^2x'  \kappa_i(\vec{x}_i)
        \ln |\vec{x}_i - \vec{x}'| \hspace{1.0cm} (i=1,2)
\end{equation}
where
\begin{equation}
\kappa_i(\vec{x}_i) \equiv \frac{\Sigma_i(\xi_i \vec{x}')}
    {\Sigma_{{\rm cr},i}} \hspace{1.0cm} (i=1,2).
\end{equation}
Here $\Sigma_i(\xi_i \vec{x}')$ are physical surface mass densities, and 
$\Sigma_{{\rm cr},i}$ are critical surface mass densities defined by
\begin{equation}
\Sigma_{{\rm cr},i}^{-1} \equiv \frac{4 \pi G}{c^2} \frac{D_i D_{is}}{D_s},
\end{equation}
similarly to the single plane lens case.

The inverse magnification matrix [${\mathcal{M}}^{-1}$] of lensing is found 
from equations (3) and (4) to be
\begin{eqnarray}
[{\mathcal{M}}^{-1}] & = &  
 \left[\frac{\partial \vec{x}_s}{\partial \vec{x}_1}\right] 
     \nonumber \\
   & = & \cal{I} - \left[ \begin{array}{cc}
                              \psi_{1, xx} & \psi_{1, xy} \\
                              \psi_{1, yx} & \psi_{1, yy} 
                             \end{array} \right] 
    -  \left[ \begin{array}{cc}
                     \psi_{2, xx} & \psi_{2, xy} \\
                     \psi_{2, yx} & \psi_{2, yy} 
                     \end{array} \right]      \nonumber \\
   &  & + \beta_{12}   \left[ \begin{array}{cc}
                              \psi_{1, xx} & \psi_{1, xy} \\
                              \psi_{1, yx} & \psi_{1, yy} 
                             \end{array} \right]  
                      \left[ \begin{array}{cc}
                     \psi_{2, xx} & \psi_{2, xy} \\
                     \psi_{2, yx} & \psi_{2, yy} 
                     \end{array} \right], 
 \end{eqnarray}
where $\cal{I}$ is a unit matrix.
Equation (10) has a cross term proportional to the distance 
between the two lens planes which would not exist for lensing by two 
deflectors in the same lens plane. 

The time delay for the light ray following a deflected light path relative to
the undeflected path from the source to the observer is given by the sum of 
delays for the two planes, i.e.,
\begin{eqnarray}
t & = & \frac{1 + z_1}{c} \frac{\xi_0^2 D_2}{D_1 D_{12}}
    \left[ \frac{1}{2} (\vec{x}_1 - \vec{x}_2)^2 - 
     \beta_{12} \psi_1(\vec{x}_1) \right]  \nonumber \\
  &  & + \frac{1 + z_2}{c} \frac{\xi_0^2 D_2 D_s}{D_1^2 D_{2s}}
    \left[ \frac{1}{2} (\vec{x}_2 - \vec{x}_s)^2 - \psi_2(\vec{x}_2) \right],
\end{eqnarray}
where $z_1$ and $z_2$ are, respectively, the redshifts of the foreground and
background deflectors. 

\section{Application to B2114+022}
\subsection{Observed Properties of B2114+022}

A full description of observed properties of B2114+022 is given in A00.
Below we only summarize the main properties and the observational constraints
to be used in lens modelling. A field of B2114+022 can be found in Figure~2,
which shows the geometric arrangements of the radio components and the galaxies
based on present absolute and relative astrometries (Table~4, A00).
\begin{figure*}
\begin{center}
\setlength{\unitlength}{1cm}
\begin{picture}(17,17)(0,0)
\put(0.,0.){\includegraphics{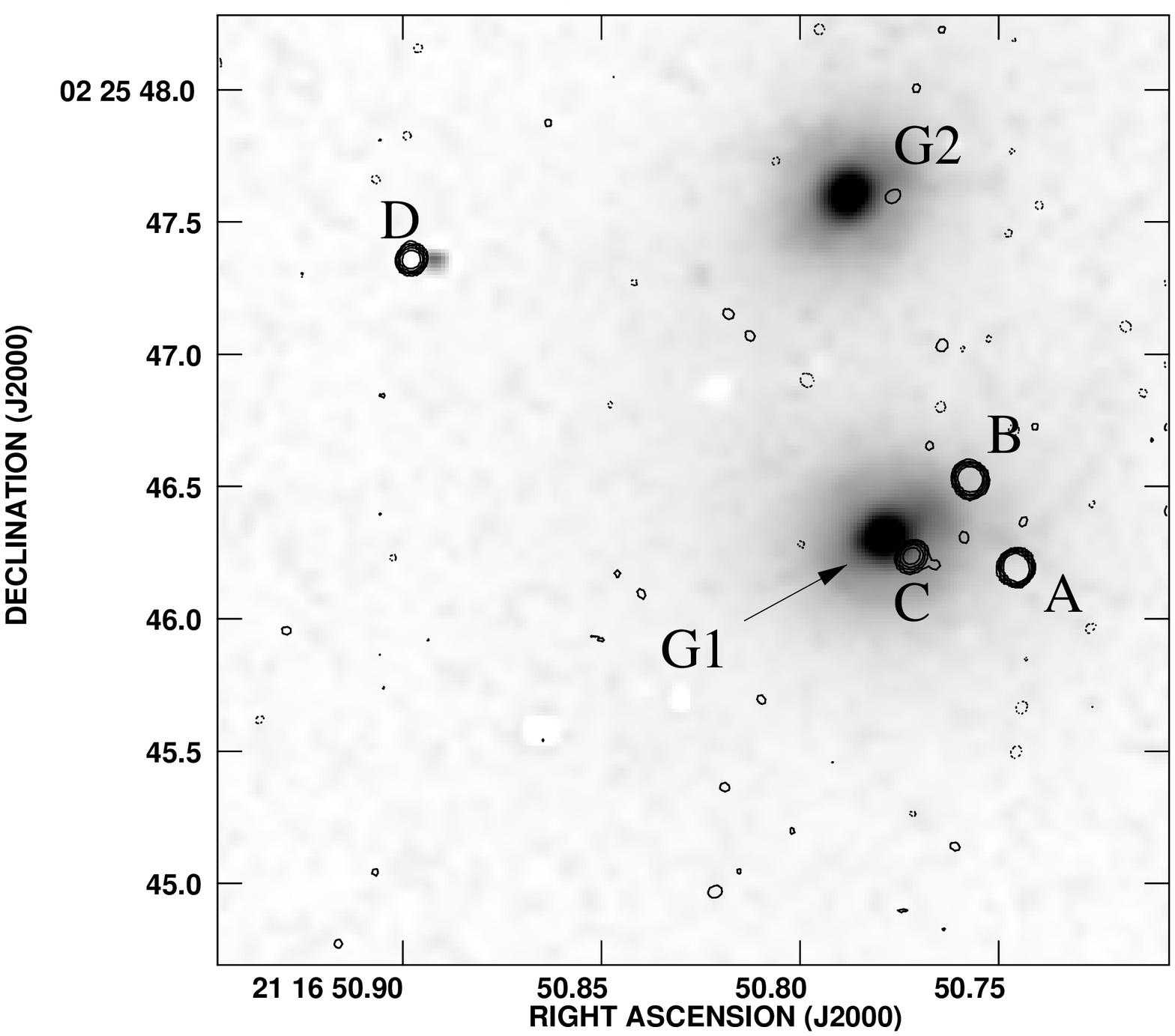}}
\end{picture}
\caption{
The observed geometric arrangement of the radio sources with respect to the
two optical galaxies in B2114+022 (reproduced from A00).
}
\label{}
\end{center}
\end{figure*}

\subsubsection{Radio Sources in a Unique Configuration}

B2114+022 is one of few JVAS/CLASS systems whose astrophysical origins 
have not been well understood to date. As shown in Figure~2,
there are four radio sources located within $\approx 2\farcs6$ of each other.
The widest separated components A and D in B2114+022 are
similar in their radio properties; they are both compact and have similar 
radio spectra over a frequency range from $\nu$~=~1.5--15~GHz, while the other 
two components found close to G1's optical centre and component A are more 
resolved and have spectral peaks at higher frequencies (Figures 2 and 3, A00). 
Two optical galaxies (see section~3.1.2) are found in the field with 
components A, B and C at the same side while only component D is at
the opposite side.

The geometric arrangements of the galaxies with respect to the radio
components and the radio imaging and spectral similarities between components
A and D hint that components A and D may be counter images of a higher
redshift radio source lensed by the potentials of the galaxies.
Based solely on the geometry of the system, it would be possible that
three or all four radio components are lensed images of the same source.
In such scenarios, however, the different radio imaging and spectral properties
of components B and C compared to those of components A and D need to be 
explained via modifications of images through the passage of the galactic
media of G1. Since such an explanation lacks observational evidence
at present, modelling components B and C
as lensed images is not motivated (at present); it will, however,
be worthwhile to revisit the issue in the future with better radio data 
and/or alternative interpretations of the present data (see section~4). 
The lensing hypothesis for components A and D is
further supported using the following simple lensing analysis. 
If the unknown redshift of component A is somewhat higher than that of G1 
($z_1 = 0.3157$) and G1 has a moderate velocity dispersion, it is required
that component A has counter image(s) because its impact parameter from G1's
centre is smaller than an Einstein ring radius using a singular isothermal
sphere (SIS) model for G1. For example, for a source redshift $z_s \ga 1$ 
and a line-of-sight velocity dispersion $\sigma_v \ga 185$ km s$^{-1}$,
an SIS Einstein ring radius is larger than
the impact parameter of component A at G1 ($\approx 0\farcs56$). 
However, the flux ratio between components A and D is uncommon in
single-plane double lenses, namely that component A (which is closer to G1)
is $\approx 3$ times brighter. This unusual flux ratio
between components A and D should then be attributed to the combined effect of
the two galaxies found in the field if they are lensed images.

Unlike components A and D,
components B and C are difficult to interpret in several ways (see A00 for
further discussion). First, components B and C are at the same side of 
G1 (with a $3\sigma$ astrometric significance), which is difficult
to reconcile with a possibility that they are double radio ejections from G1. 
Second, they have more extended image structures while their spectra 
peak at higher frequencies compared with the more compact components A and D. 
Finally, both components B and C are within 0\farcs5 of G1,
which is smaller than an Einstein ring radius of G1 for a moderate velocity
dispersion of G1 and an intermediate source redshift (see above), 
and the angular separation between components B and C
is $0\farcs36$, which is 7 times smaller than the 
separation between components A and D, i.e. $2\farcs56$. This last
point from the geometry of B2114+022 virtually rules out a possibility
that components B and C are lensed images of an independent source due to
the same potentials which are supposed to give rise to components A and D,
although it would not be inconsistent with a possibility that they are 
counter images of components A and D.

Deep HST optical/near infrared observations of B2114+022 have not resulted in
any detection of optical counterparts of the radio components down to $I = 25$
(WFPC2, F814W) and $H = 23$ (NICMOS, F160W). This may not be surprising for
components B and C because of their proximity to G1's centre. The apparent
extreme faintness of components A and D at optical wavelengths may indicate 
that their source is a high redshift and/or an intrinsically faint object,
such as a low luminosity radio galaxy. 

Table~1 summarizes A00's relative positions of radio components B, C and D and
the two galaxies with respect to component A, and the flux density ratios of 
the radio components. These flux density ratios do not include A00's 1.6~GHz 
data (Table~3 of A00), at which the spectra of components B and C are already 
turned over while those of components A and D start to turn over (see Figure~3
of A00). The flux ratio of D/A most relevant for this study is not 
significantly affected if 1.6~GHz data are included. 
\begin{table}
\caption{Relative positions of radio and optical sources and radio flux
density ratios. Observations used are as follows.
(1) Radio component positions: MERLIN 1.6 GHz, VLBA 5.0 GHz,
 MERLIN+EVN 1.6 GHz. (2) Radio component flux density ratios: VLA 8.4 GHz,
 15 GHz, MERLIN 5 GHz. (3) Relative positions of G1 with respect to
 radio component A: NOT I band + VLA 8.4 GHz. (4) Relative positions of
G2 with respect to G1: NICMOS H band.}
\begin{tabular}{clll} \hline
Component & $\Delta\alpha$ (arcsec) &  $\Delta\delta$ (arcsec) & 
                               $R_{f_{\nu}}$  \\ \hline
 A   & $\equiv 0$  & $\equiv 0$           & $\equiv 1$   \\
 B   & $0.175\pm 0.001$  & $0.333\pm 0.001$  & $0.90\pm0.03$     \\
 C   & $0.397\pm 0.001$  & $0.043\pm 0.001$  & $0.28\pm0.02$     \\
 D   & $2.286\pm 0.001$  & $1.158\pm 0.002$  & $0.32\pm0.02$    \\
 G1  & $0.555\pm 0.1$  & $0.04\pm 0.1$  & ---    \\
 G2  & $0.145\pm 0.006$ $^a$  & $1.296\pm 0.006$ $^a$  & ---     \\ \hline
\end{tabular}

$^a$ Relative positions of G2 with respect to G1
\end{table}

\subsubsection{Two Close Galaxies with Different Redshifts}

The two galaxies found close to the radio components have
different redshifts, namely $z_1 = 0.3157$ for G1 and $z_2 = 0.5883$ for G2,
making them the first possible two plane lens. G1 is nearly on the line
joining components A and D while G2 is misaligned with the line.
Within the framework of single plane lensing, this would seem to
suggest that G1 is responsible for most of the lensing while G2
provides only a second-order effect in a lensing hypothesis. However, this is 
not quite so for two reasons. First, the observed position of G2 is a deflected
position due to G1. Second, light rays from the source were deflected by G2 
before they were deflected by G1 finally forming the observed images. In fact,
the undeflected source position can be closer to (the undeflected) G2 than G1
in a reasonable two plane lens model (section~3.2.2 and Figure~6). 
\begin{figure*}
\begin{center}
\setlength{\unitlength}{1cm}
\begin{picture}(17,13)(0,0)
\put(0.,12.5){\includegraphics{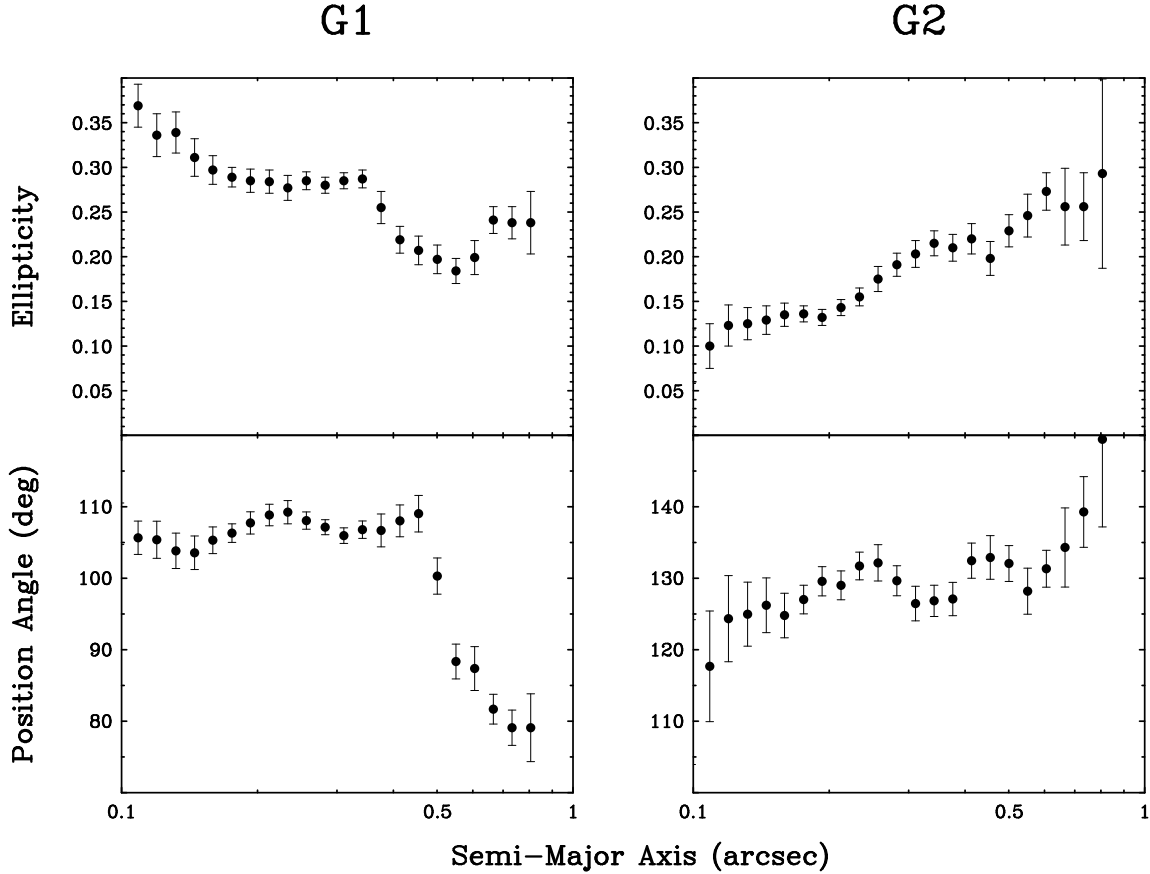}}
\end{picture}
\caption{
Ellipticity and position angle as a function of semi-major axis
(in arcseconds) for the foreground (G1) and background (G2) galaxies
derived from HST WFPC2 I band (F814W) images of the galaxies.
}
\label{}
\end{center}
\end{figure*}

HST WFPC2 V (F555W) and I (F814W) band and NICMOS H (F160W) band observations 
of the galaxies reveal that both galaxies are moderately elliptical with
fitted ellipticities and position angles somewhat fluctuating as a function
of semi-major axis (see Figure~3). Application of a K-correction to A00's 
measured magnitudes of G1 and G2 in the above wave-bands, by assuming 
$S_{\nu} \sim \nu^{-2}$, gives an estimate of the apparent luminosity ratio of 
$L$(G2)/$L$(G1)~$\sim 3$ (N. Jackson; personal communication). 
However, since G2's apparent luminosity was 
magnified due to G1's potential, we estimate a true luminosity ratio of
$L$(G2)/$L$(G1) $\sim 2$ after correcting for a magnification of $\sim 1.4$
for G2 (see Figure~5). For similar mass-to-light ratios for the two
galaxies, this luminosity ratio would imply that G2 is more massive than G1.

\subsection{Lens Modelling}

In this section we investigate possible lens models of B2114+022. One goal
of doing so is to test lensing hypotheses for this system, 
in particular the possibility of reproducing components A and
D using an astrophysically plausible model. If one can find a successful
lens model, the lensing hypothesis will be strengthened. This is particularly
important for this system since optical spectroscopy of the radio components
appears to be extremely difficult due to extreme faintness of the radio sources
at optical wavelengths and relatively large contamination from the galaxies
(see section~3.1.1). Another goal is to study two plane lensing,
particularly by making comparisons of the lensing properties of
single plane lens models and two plane lens models for B2114+022.

In section~3.2.1, we first consider a popular single plane lens model,
namely a power-law ellipsoid plus a shear, in which the shear term is intended
to account for the lensing effect due to G2. In section~3.2.2, we consider two
plane lensing by explicitly including G2's potential at the observed redshift.
For each galaxy, we adopt a power-law mass model which includes an isothermal
model as a special case, i.e., surface mass density of the form,
\begin{equation}
\Sigma(\xi,\theta)=\frac{\Sigma_0}{\left\{1+\left(\frac{\xi}{\xi_c}\right)^2
[1 + e \cos 2(\theta-\theta_0)] \right\}^{(\nu-1)/2}},
\end{equation}
where $\nu = 2$ corresponds to an isothermal radial index, 
$\xi \equiv \sqrt{\xi_x^2 + \xi_y^2}$, parameter $e (> 0)$ is related to the
ellipticity via $\epsilon = 1 - \frac{\xi_{\mbox{\scriptsize min}}}
{\xi_{\mbox{\scriptsize max}}} = 1 - \sqrt{\frac{1 - e}{1 + e}}$, 
$\theta_0$ is the standard position angle (P.A., north through east), and
$\xi_c$ is a core radius.
We calculate lens models in a cosmology with $\Omega_m = 1$, 
$\Omega_\Lambda = 0$, and $H_0 = 60 h_{60}$ km s$^{-1}$ Mpc$^{-1}$.
In this cosmology, one arcsecond corresponds to $4.7h_{60}^{-1}$ kpc and
$6.3h_{60}^{-1}$ kpc on the foreground and background lens planes respectively.
Model parameters and predictions are dependent on the unknown 
source redshift ($z_s$) especially for $z_s \la 1.5$; however, for 
$z_s \ga 2$, the dependence is little. All parameters and predictions given
below are for $z_s = 3$.

As was pointed out in section~3.1.1, components A and D would be consistent 
with a lensing hypothesis while the origin of components B and C is not
obvious, at best. We thus use the observed positions of components A and D and
their flux ratio as model constraints in this study. For a single plane lens 
model, G1's positions provide two additional constraints, while for a 
two-plane lens model, both galaxies provide four additional constraints.
In total, there are 5 and 7 (direct) constraints for the single-plane and
two-plane lensing cases, respectively. 

The small number of constraints is a major difficulty in investigating lens 
models for B2114+022. For example, a mass model of the form given by equation
(12) with fixed radial index and core radius even without a shear term has as
many free parameters (i.e.\ $\epsilon$, $\theta_0$, $\Sigma_0$, and the two
coordinates of the source on the source plane)
as observational constraints. Fortunately, however, 
there are additional pieces of information on the galaxies from HST imaging
which could not be quantitatively used in lens modelling, nevertheless can
provide crucial tests of lens models as regards astrophysical plausibility. 
For this purpose, the observed position angles and ellipticities and inferred 
luminosity ratio of the galaxies are (or can be) useful. First of all,
a mass distribution significantly misaligned with the observed light 
distribution is unlikely since the study of an ensemble of gravitational 
lenses shows a general trend that the inferred mass distributions are 
aligned with the observed light distributions within $\sim 10\degr$ (Keeton, 
Kochanek \& Falco 1999) at least for the lensing galaxies which are relatively
isolated or in low density environments. Analyses of X-ray isophotes of 
three isolated early-type galaxies NGC 720, 1332 and 3923 [see the review
by Buote \& Canizares (1997) and references therein] show the general 
alignment between the mass and light, except that in NGC 720, for outer
parts of the galaxy ($>R_e$) the mass and light are misaligned by 
$\sim 30\degr$. Analyses of the kinematics of polar rings in polar-ring
galaxies also find the general alignment between the mass and light 
(Arnaboldi et al.\ 1993, Sackett et al.\ 1994, Sackett \& Pogge 1995).
Second, while the shape of dark matter distribution in galaxies is poorly
known at present (see the review by Sackett 1999), the above X-ray studies by 
Buote \& Canizares (1997) report that the inferred mass ellipticities are 
similar to or somewhat higher than the observed optical light ellipticities.
The individual modelling of lensing galaxies by Keeton et al.\ (1999) does not
find any clear correlation between the light and inferred mass ellipticities;
mass ellipticities are higher than light ellipticities for some lensing 
galaxies while it is the opposite for others. Given the present limited 
knowledge of the dark matter shape in galaxies, the observed light 
ellipticities of
G1 and G2 can only provide a very limited test of lens models. For example, 
a disk-like mass shape for either G1 or G2 is unlikely given that the 
observed light distributions are only moderately elliptical. 
Finally, the luminosity ratio of the two galaxies can provide some 
information on the mass ratio of the two galaxies. 

\subsubsection{Test of Single Plane Lens Model: Ellipsoid plus Shear}

To model B2114+022 with a single-plane lens, 
we need to incorporate G2's lensing effect, which to first order, can be
modeled as a shear term. Its deflection is given by
\begin{equation}
\vec{\alpha}_{\gamma}(\vec{x}) = \gamma \left( \begin{array}{rr}
       \cos 2\theta_{\gamma} & \sin 2\theta_{\gamma} \\
       \sin 2\theta_{\gamma} & - \cos 2\theta_{\gamma} \\
      \end{array} \right) \vec{x},
\end{equation}
where $\gamma$ and $\theta_{\gamma}$ are the shear strength and P.A. 
respectively.
Since the shear term is intended to account for G2's lensing effect, we expect
that the shear points to G2's direction from G1,\footnote{In single plane 
lensing, we assume that the two galaxies are projected onto a single plane. 
A choice on the redshift of the plane is irrelevant here since we do not know 
the source redshift. The lensing effect of G1 changes G2's apparent position
relative to G1 primarily in the radial direction; it affects the position
only slightly tangentially.} which roughly corresponds to 
$\theta_{\gamma}\sim 0\degr$. A shear oriented in a direction significantly
different from this presumed direction would be difficult to interpret within
the framework of single plane lensing. We thus fix 
$\theta_{\gamma} =0 \degr$ and investigate the effect of
positive shear strength on the fitted galaxy parameter values. 

Although we expect that a non-zero shear is necessary to fit the data (i.e.\
the relative positions of G1 and component D with respect to component A and
the flux ratio between components A and D) with more suitable parameter 
values, we first consider a zero shear case to merely keep records of the 
determined parameter values to be compared with those for the non-zero
shear case below. We find that for the isothermal ($\nu =2$) galaxy profile, 
the data can be perfectly fit provided that $\epsilon = 0.77$ and
$\theta_0 = 146\degr$. For a somewhat shallower profile of $\nu = 1.75$
(a profile shallower than this is excluded since a theoretical image near the
mass centre becomes brighter than a $5\sigma$ radio flux limit on any
unobserved image, i.e., 2.4\% of component D's flux; Norbury et al.\ 2000), 
the required ellipticity is reduced to $\epsilon = 0.58$. 
For this zero (external) shear case of the single plane lens model,
the required ellipticity and P.A. are, respectively, much higher than
and significantly misaligned with those of G1's observed isophotes.

As the shear strength is increased, 
the model galaxy's P.A. consistently rotates in the way that
the misalignment with G1's isophotes is reduced while the model galaxy's 
ellipticity remains relatively unchanged [specifically, it falls only 
moderately for small shear (e.g.\ $\gamma < 0.1$ for $\nu =2$) but rises 
even higher for larger shear values]. For the isothermal galaxy model 
(and $\nu=1.75$ model) considered above, for the galaxy to be aligned with
the light (choosing P.A.$_{\mbox{\scriptsize light}} = 110\degr$),
the required galaxy ellipticity and shear strength are $\epsilon = 0.90$ 
($\epsilon = 0.82$) and $\gamma = 0.40$ (the same). Thus, the high ellipticity
problem persists with the inclusion of a shear term. If one introduces an
arbitrary shear rather than the above astrophysically motivated shear within
single plane lensing framework, in other words if one allows the shear 
orientation to be arbitrary, it is possible to fit the data with likely galaxy
parameters based on the observed light distributions. For the isothermal
galaxy model (and $\nu=1.75$ model) considered above, a galaxy with 
$\epsilon=0.3$ and P.A. $= 110\degr$ can fit the data providing 
$\gamma = 0.47$ ($\gamma = 0.36$) and $\theta_{\gamma} = -28\degr$ 
($\theta_{\gamma} = -26\degr$). However, as pointed out above, 
these shear orientations are not expected for single plane lensing.

To summarize the results of single plane lens modelling of B2114+022 A,~D,
for the model galaxy allowed to be misaligned with G1's light, the minimum
required  ellipticity is $\epsilon = 0.76$ ($\epsilon = 0.57$) for $\nu = 2$ 
($\nu = 1.75$), while for the model galaxy aligned with G1's light, the model
requires $\epsilon = 0.90$ ($\epsilon = 0.82$) as well as a shear of 
$\gamma = 0.40$  for $\nu = 2$ ($\nu = 1.75$). For this latter case of a
likely mass position angle of G1, the required large shear strength and
the required very flattened projected mass density of G1 are unrealistic. 
This failure of single plane
lens models in explaining the simple lensing constraints of B2114+022 leads
us to one of the following two possibilities: Either single plane lensing
is not applicable to this system because G2's lensing effects are comparably
important or the lensing hypothesis on components A and D is doubtful.
However, the latter possibility cannot be justified unless one has applied
the more accurate theory to the system, namely two plane lensing theory.
Next we investigate two plane lens models of B2114+022.

\subsubsection{Two Plane Lens Model}
The formalism of two thin plane lensing was reviewed in section~2. Using this
formalism and adopting the mass model of equation~(12) for each of the two
galaxies at their observed positions and redshifts, we calculate the 
theoretical deflection, magnification and time delays due to the two 
deflectors using a code employing Fourier expansion techniques (Chae, 
Khersonsky \& Turnshek 1998), and then fit the observed positions and flux 
ratio of components A and D by varying model parameter values of the galaxies.
For this problem, we are fortunate to have relatively
high-quality optical data on the two galaxies. The redshifts and optical
centres of the two galaxies are directly used to constrain the lens model.
The observed light distributions (in terms of ellipticities and position
angles) of the two galaxies can be used to test astrophysical plausibility
of the model ellipticities and position angles. While the lower redshift 
galaxy G1's observed properties are intrinsic properties, the higher
redshift galaxy G2's observed properties are modified (i.e.\ lensed) 
properties of G2 due to the potential of G1. The relatively small angular
separation between the two observed galaxies ($d_{12} \approx 1\farcs30$, which
corresponds to the impact parameter for light rays from the G2 centre at the 
lensing plane of G1) ensures that the lensing of G2 by G1 is significant. In 
particular, we expect a shift in the galaxy's optical centre, stretching of
light ellipses approximately along the east-west line (i.e.\ perpendicularly to
the line joining G1 and G2) and possibly additional 
distortions of the stretched ellipses beyond a certain radius
(i.e.\ arc-like light shapes). 

Solving two plane lensing involves solving the single plane
lensing of the extended background source by the foreground deflector. 
Ultimately, the lensed light distribution of G2 can be derived by 
subtracting out G1's light distribution, and the resulting extended arc-like 
light distribution can be used as a lensing constraint on G1's potential.
However, due to our present lack of a suitable means to quantitatively deal
with this problem and limited quality of the presently available data for
this purpose, in this paper we do not fit G2's light distribution but
use it to qualitatively test astrophysical plausibility of the intrinsic
ellipticity and position angle of the G2 mass model. We know {\it a posteriori}
that for comparable masses of G1 and G2 for the observed geometric 
arrangement, the lensing effect of G1 leads to a net increase of G2's 
ellipticity by $\approx$ 0.2--0.3 along approximately east-west.
This allows us to limit possible ranges of the ellipticity and position angle
of G2's intrinsic light distribution (which can in turn provide some 
information on G2's mass distribution). Loosely speaking, an intrinsically
elliptical G2 cannot be oriented along east-west since such a light 
distribution would become more elliptical than observed due to a net
increase in the same direction caused by G1, and thus it should be roughly 
oriented along north-south. The intrinsic ellipticity of G2's light then 
depends on a stretching by G1 and the measured value of G2
image's ellipticity. For example, for a measured ellipticity of 0.1 and
an amount of stretching of 0.25, the required G2's intrinsic ellipticity is
0.15 for a P.A. of zero.

For a given radial profile ($\nu$) for each model galaxy, the mass 
distribution is determined by parameters $\Sigma_0$, $\xi_c$, $\epsilon$ and 
$\theta_0$ [equation (12)]. However, since neither $\Sigma_0$ nor $\xi_c$ is 
sensitive to the lensing properties of the model (other than a relationship 
between $\xi_c$ and relative magnification of a theoretical image near the 
mass centre), we use an Einstein ring radius ($\xi_E$) which is directly
related to the deflection scale of the model galaxy and insensitive to the 
choice of either $\xi_c$ or $\Sigma_0$.\footnote{For a $\xi_E$, fixing either
$\xi_c$ or $\Sigma_0$ determines the other.}
We use a $\xi_E$ determined from 
\begin{equation}
\frac{\xi_E^2}{\xi_c^2} = \frac{2 \kappa_0}{3 -\nu}\left[
 \left(1 + \frac{\xi_E^2}{\xi_c^2}\right)^{\frac{3-\nu}{2}} - 1 \right]
\end{equation}
for the mass model of equation (12). For a single-galaxy lens, parameter
$\xi_E$ is well constrained regardless of the choice of lens model parameters.
For the two-galaxy model under consideration, the lensing effect is
the combined effect due to the two deflectors. Thus, we do not expect
individual Einstein ring radii of the galaxies to be well constrained
for a two-galaxy lens system, while we may expect the `sum' of them to be.
A parameter controlling the relative size of two Einstein ring radii 
(or masses enclosed within them) is a fundamental parameter of interest
in a two-galaxy system which we intend to constrain.  
In this study, we define a total Einstein ring radius 
$\xi_E^{\mbox{\scriptsize tot}} \equiv \xi_E^{(1)} + \xi_E^{(2)}$ and 
a parameter
\begin{equation}
f_2^{R} = \frac{\xi_E^{(2)}}{\xi_E^{\mbox{\scriptsize tot}}},
\end{equation}
which controls the relative size of Einstein ring radii of the two galaxies.
We also define a parameter
\begin{equation}
f_2^M = \frac{M_E^{(2)}}
        {M_E^{\mbox{\scriptsize tot}}},
\end{equation}
where $M_E^{\mbox{\scriptsize tot}} = M_E^{(1)} + M_E^{(2)}$
and $M_E^{(i)}$ ($i=1,2$) are the masses enclosed within the
Einstein ring radii of G1 and G2 respectively. 

We first consider isothermal profiles $\nu^{(i)} = 2$ ($i=1,2$ where labels 
1, 2 denote G1, G2 respectively hereafter). Out of the six remaining 
parameters of the two galaxies, some of the parameters are {\it a posteriori}
trivial or confined within relatively small ranges. As was discussed above,
we expect G2's mass distribution to be oriented approximately 
north-south. Remarkably, we find that in order to fit the data G2's P.A. 
($\theta_0^{(2)}$) has to be approximately within a quadrant around 
north. We fix G2's P.A. at $\theta_0^{(2)} = +10\degr$ since at other angles 
the required ellipticity of G2 is higher and we do not expect a high intrinsic
ellipticity for G2 (see the discussion above). As is always the case in
lensing, an appropriately defined total mass (or, equivalently deflection 
scale) for lensing is expected to be well constrained. For the two-galaxy 
model under consideration, parameter $\xi_E^{\mbox{\scriptsize tot}}$
is confined within a relatively small range. However, parameter $f_2^{R}$ 
is not well constrained by (at least present) lensing constraints. 
We fix $f_2^{R}$ for each model, and increment its value between 0.3 
and 0.5 which were chosen {\it a posteriori}. We expect 
the position angle of G1's mass model ($\theta_0^{(1)}$) to be similar
to that of G1's observed light distribution, i.e.
$100\degr \la {\mbox{P.A.}}_{\mbox{\scriptsize light}}^{(1)}  \la 110\degr$ 
for $0''.1 \la r_{\mbox{\scriptsize max}} \la 0''.5$ (Figure~3). 
We find, however, that the fitted value of $\theta_0^{(1)}$ 
for a large region of parameter space tends to be 
more north-south oriented than P.A.$_{\mbox{\scriptsize light}}^{(1)}$. 
We fix $\theta_0^{(1)}$ for each model and increment its value in the range
between $100\degr$ and $150\degr$. 

With the above prescriptions, we determine
the other parameters (i.e., $\epsilon^{(1)}$, $\epsilon^{(2)}$ and 
$\xi_E^{\mbox{\scriptsize tot}}$) for each model by varying them 
simultaneously. In this way we calculate a grid of models covering realistic
ranges of model parameters. Due to the small number of observational 
constraints, each model gives a perfect fit to the data. Parameters 
$\xi_E^{\mbox{\scriptsize tot}}$ and $\theta_0^{(2)}$ being trivial,
parameters of our primary interest are $\epsilon^{(1)}$, $\theta_0^{(1)}$, 
$\epsilon^{(2)}$ and $f_2^R$.
 We find that for most models in the grid $\epsilon^{(1)}$ is confined
within a relatively small range of $0.3 \la \epsilon^{(1)} \la 0.4$.
Thus, the required extraordinarily high ellipticity for G1 encountered in
the single plane lens model (considered in section~3.2.1) is not necessary
when G2's potential is included at its observed redshift. Correlations among
the other three parameters $\theta_0^{(1)}$, $f_2^R$ and $\epsilon^{(2)}$
can be found in Figure~4a which also shows the parameter $f_2^M$. 
\begin{figure*}
\begin{center}
\setlength{\unitlength}{1cm}
\begin{picture}(22,15)(0,0)
\put(-1.,14.5){\includegraphics{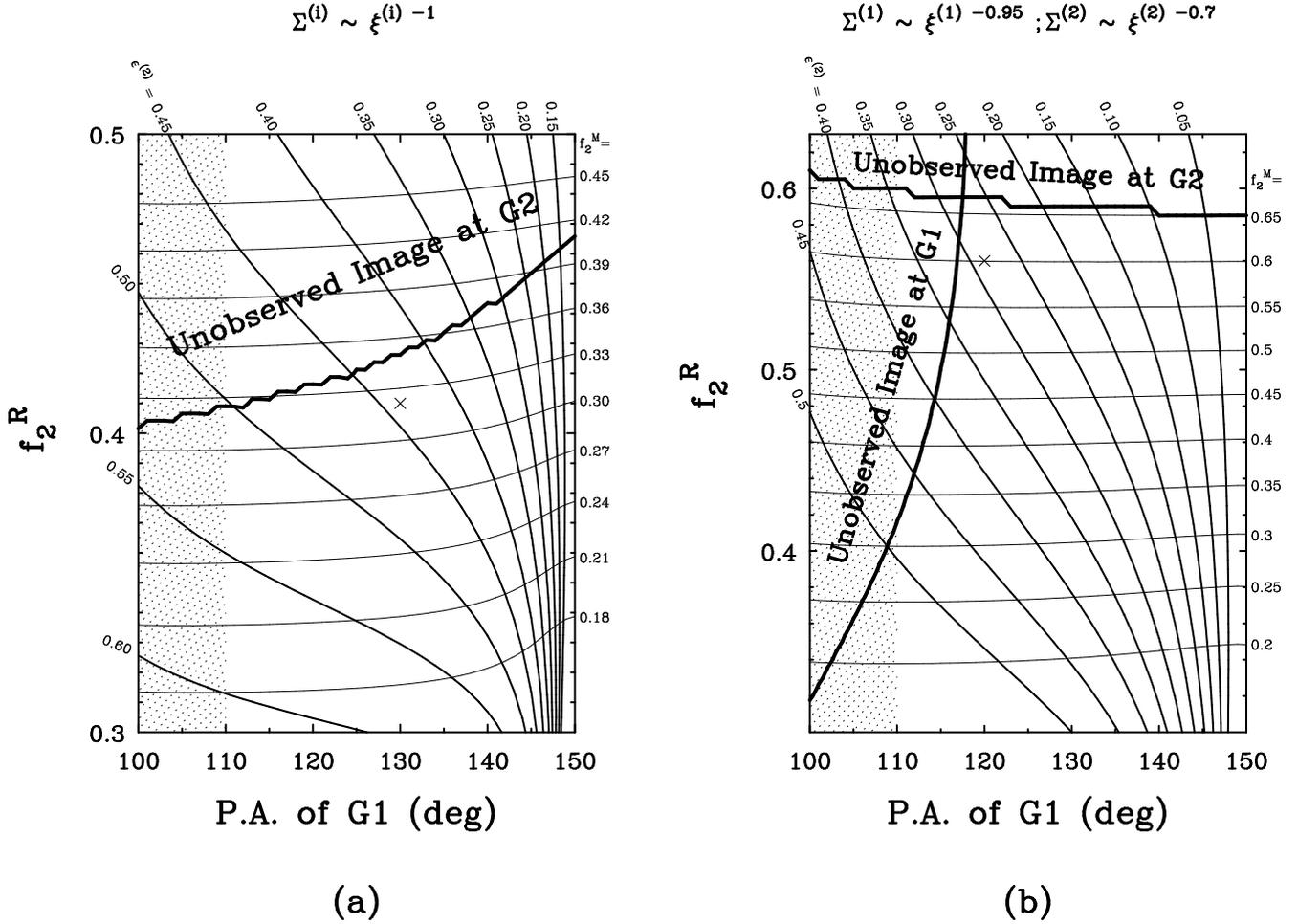}}
\end{picture}
\caption{
(a) Correlations among parameters $\theta_0^{(1)}$
(G1's P.A.), $f_2^R$ [equation (15)] and $\epsilon^{(2)}$ (G2's ellipticity)
for two isothermal galaxies model.
Parameter $f_2^M$ defined by equation (16) is also shown for a reference
to the mass ratio between the two galaxies.
The hatched region is the observed range of position angle.
The parameter space above the thick solid line is
ruled out due to predicted additional bright images near G2
that are not observed. A cross indicates an example model whose
parameters and predictions are given in Table~2.
(b) Similar correlations among the same parameters as in (a) for the mass
profiles $\nu^{(1)} = 1.95$ and $\nu^{(2)} = 1.7$ for G1 and G2 respectively.
Compared with the two isothermal mass profiles model in (a), 
for these shallower mass profiles the upper excluded region due to additional
images near G2 is pushed upward while a region left of the thick vertical
line is now excluded due to a predicted bright image near G1.  
}
\label{}
\end{center}
\end{figure*}
An upper limit on $f_2^R$ (equivalently $f_2^M$) for each $\theta_0^{(1)}$ 
is set by the requirement that any predicted image near G2's centre
should be fainter than the observational flux limit on any unobserved image 
(i.e.\ 2.4\% of component D flux density). Thus, G2 can only be
somewhat less massive than G1 based on the two isothermal galaxies model. 
Combined with the observationally inferred luminosity ratio between the
two galaxies of $L$(G2)/$L$(G1) $\sim 2$ (section~3.1.2), the upper limits on 
$f_2^M$ (Figure~4a) would imply that the mass-to-light ($M/L$) ratio
of G1 is higher than that of G2 at least by a factor of $\sim 4$.
As a consequence of this relatively large implied mass of G1 in the isothermal
galaxies model, the model predicts relatively large stretching and distortion 
of G2's light distribution. Table~2 gives parameters 
and predictions of an example two isothermal 
galaxies model marked by a cross in Figure~4a (Model~1). 
\begin{table*}
\caption{Example two-plane lens model parameters and predictions.
For Model~1, both G1 and G2 have isothermal profiles, while for Model~2,
G1 and G2 have different and shallower-than-isothermal profiles. The parameter
values given are for the models marked by a cross in Figure~4a,b. The parameter
ranges given in parentheses are for models with 
$\epsilon^{(i)} < 0.5$ ($i = 1,2$) and 
$|\theta_0^{(1)} - {\mbox{P.A.}}_{\mbox{\scriptsize light}}^{(1)}| < 30\degr$.}
\begin{tabular}{llll} \hline
Parameter & Description  &   Model 1 &   Model 2 \\ \hline
$\xi_E^{\mbox{\scriptsize tot}}$ ($h_{60}^{-1}$ kpc) &
Sum of Einstein radii of G1 \& G2 &  7.11 (6.6--7.2)  &  6.85 (5.7--7.7)\\
$f_2^R$  & See equation (15).  &  0.41 (0.33--0.44) &  0.56 ( $<$ 0.6)  \\
$\nu^{(1)}$ & Radial index of G1 & 2   &  1.95  \\
$\nu^{(2)}$ & Radial index of G2  & 2  &  1.7  \\
$\epsilon^{(1)}$ & Ellipticity of G1 &  0.32 (0.30--0.44) 
                               &  0.37 (0.14--0.5) \\
$\epsilon^{(2)}$  & Ellipticity of G2 &  0.44 (0.32--0.5)  
                     &  0.29 (0.09--0.5) \\
$\theta_0^{(1)}$ (deg) & P.A. of G1 &  130 (111--140) & 120 (108--140)  \\
$\theta_0^{(2)}$ (deg) & P.A. of G2 &  10  &  10  \\ \hline
Prediction   &     &     \\ \hline
$\xi_c^{(1)}$ ($h_{60}^{-1}$ kpc) & Core radius of G1 
          &  $<0.073$   & $<0.010$  \\
$\xi_c^{(2)}$ ($h_{60}^{-1}$ kpc) & Core radius of G2 & no limit & no limit \\
$M_E^{\mbox{\scriptsize tot}}$ ($10^{11} M_{\odot}$) & 
   Sum of Einstein masses of G1 \& G2 
  & 1.92 (1.8--2.0)  &   1.70 (1.4--2.4) \\
$f_2^M$   & See equation (16). &  0.32 (0.19--0.38)  &  0.60 ( $<$ 0.67) \\
$t_{AD}$  ($h_{60}^{-1}$ days) & Time for A delaying D &  86.3 (83.--89.) 
  &  66.4 (47.--81.)  \\
${\mathcal{M}}_A + {\mathcal{M}}_D$ & Sum of magnifications for A \& D 
         &  10.3 (8.8--11.) &  16.6 (10.6--30.5) \\ \hline
\end{tabular}
\end{table*}
The ranges given in the table are only for models with 
$\epsilon^{(i)} < 0.5$ ($i = 1,2$) and 
$|\theta_0^{(1)} - {\mbox{P.A.}}_{\mbox{\scriptsize light}}^{(1)}| < 30\degr$
in the grid. Predicted lensing of G2's light distribution by G1 is shown in 
Figure~5: G1's lensing on G2 leads to a stretching of G2's light
distribution by $\Delta \epsilon \approx +0.3$ approximately along east-west.
For an observed light ellipticity of 
$\epsilon_{\mbox{\scriptsize obs}}^{(2)} \approx 0.15$ (Figure~3),
a stretching of $\Delta \epsilon \approx +0.3$ implies an intrinsic
ellipticity of $\epsilon_{\mbox{\scriptsize int}}^{(2)} \approx -0.15$ 
along east-west (i.e.\ $\epsilon_{\mbox{\scriptsize int}}^{(2)} \approx +0.15$ 
along north-south). From the two isothermal galaxies model grid of Figure~4a,
we find that the required mass ellipticity of G2 is much higher than the
inferred intrinsic light ellipticity of G2, especially for models in which
G1's light and mass are aligned (the hatched region in Figure~4a) [while
the required mass ellipticity of G1 is similar to or slightly higher than 
the light ellipticity of G1]. This could be taken as a model-dependent evidence
for a significantly flattened dark matter halo for G2, under the condition that
G2's radial mass profile is isothermal and G1's light and mass are aligned
[this latter condition is likely to be valid (see above in section~3.2)].
Conversely, the required high ellipticity of G2 could be taken as an argument
for a shallower-than-isothermal profile for G2, which we consider below. 
It is also worth emphasizing that the two isothermal galaxies model imply a 
large difference between $M/L$ ratios of the two galaxies (at least by a 
factor of $\sim 4$; see above). 

\begin{figure*}
\begin{center}
\setlength{\unitlength}{1cm}
\begin{picture}(16,12)(0,0)
\put(-1.1,12.){\includegraphics{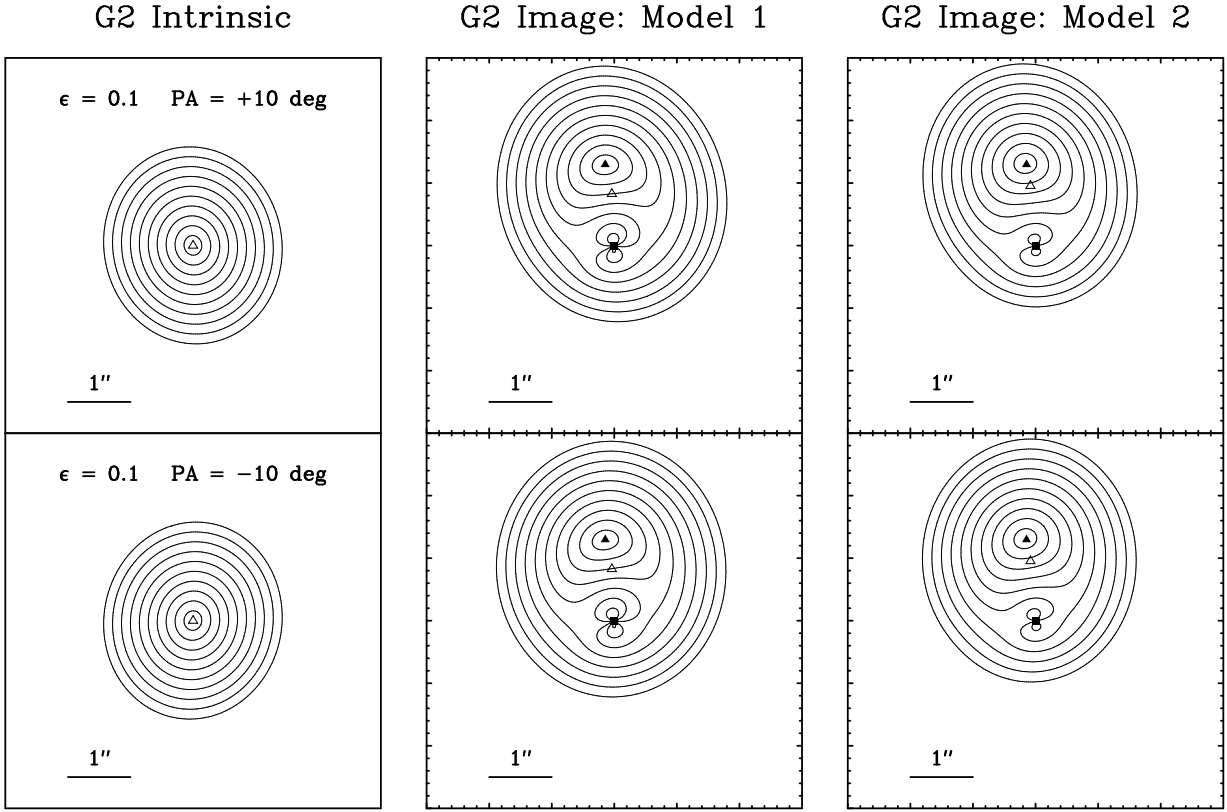}}
\end{picture}
\caption{
Predicted shift, stretching and distortion of the background galaxy's (G2)
isophotes induced by the foreground galaxy (G1). An intrinsic G2's ellipticity
of 0.1 for two position angles P.A. $= +10\degr$ (aligned with the mass model 
P.A.) and P.A. $= -10\degr$ are considered (left panels). The 10 ellipses 
correspond to semi-major axes of 1, 2, ..., 10 $h_{60}^{-1}$kpc at 
$z_2 = 0.5883$. Middle panels and right panels are predicted images for 
Model~1 and Model~2 (Table~2 and Figure~4) respectively. Notice that the 
inner several kpc regions of G2 have
P.A.s nearly orthogonal to the intrinsic P.A.s in these models.
The filled squares and triangles denote observed G1 and G2 positions while
the open triangles denote the model-predicted intrinsic position of G2.
}
\label{}
\end{center}
\end{figure*}

The relatively high ellipticity of G2 required in the two isothermal galaxies
model can be significantly reduced by making the radial profile of G2 somewhat 
shallower than isothermal. In this case the required ellipticity for G1 of an
isothermal profile slightly increases. However, this increase
in G1's ellipticity can be avoided by making the radial profile of G1
slightly shallower than isothermal. Similarly to the case in the single
plane lens model considered in section 3.2.1, for a significantly 
shallower-than-isothermal profile of G1 a theoretical image near G1's centre
becomes brighter than the observational limit, 
insensitive to the core radius of G1.
A model with $\nu^{(1)} = 1.95$ and $\nu^{(2)} = 1.7$ was calculated and 
relationships among parameters $\theta_0^{(1)}$, $f_2^R$ and
$\epsilon^{(2)}$ can be found in Figure~4b.  In this model, parameter
$f_2^M$ [equation (16)] covers a wider range compared with the two
isothermal galaxies model, including values implying G2 more massive than G1.
In particular, within the upper limits of $f_2^M$, G2 could be twice as 
massive as G1 implying that both galaxies have similar $M/L$ ratios.
As Figure~4b shows, with the chosen radial profiles for the two galaxies, 
one can find a model in which model position angles and ellipticities for both
G1 and G2 are not too different from those for the observed light 
distributions. The parameter values for such a model (marked by a cross in 
Figure~4b) are given in Table~2 (Model~2). Predicted lensing of G2's light
distribution by G1 is shown in Figure~5. In this model the amount of predicted
stretching for G2 is $\Delta \epsilon \approx 0.2$, which implies an intrinsic
light ellipticity of $\epsilon_{\mbox{\scriptsize int}}^{(2)} \approx 0.05$
along north-south for $\epsilon_{\mbox{\scriptsize obs}}^{(2)} \approx 0.15$.
Compared with the predictions of Model~1 (Figure~4a), 
Model~2 predicts somewhat weaker distortions.
In Model~2, the mass ellipticities are relatively low but slightly higher 
than their observed/inferred intrinsic light ellipticities and G1's 
mass distribution is misaligned with G1 light ellipses by only $\sim 10\degr$.
The critical curves and caustics for this model are shown in Figure~6.
\begin{figure*}
\begin{center}
\setlength{\unitlength}{1cm}
\begin{picture}(17,17)(0,0)
\put(-0.5,-3.){\includegraphics{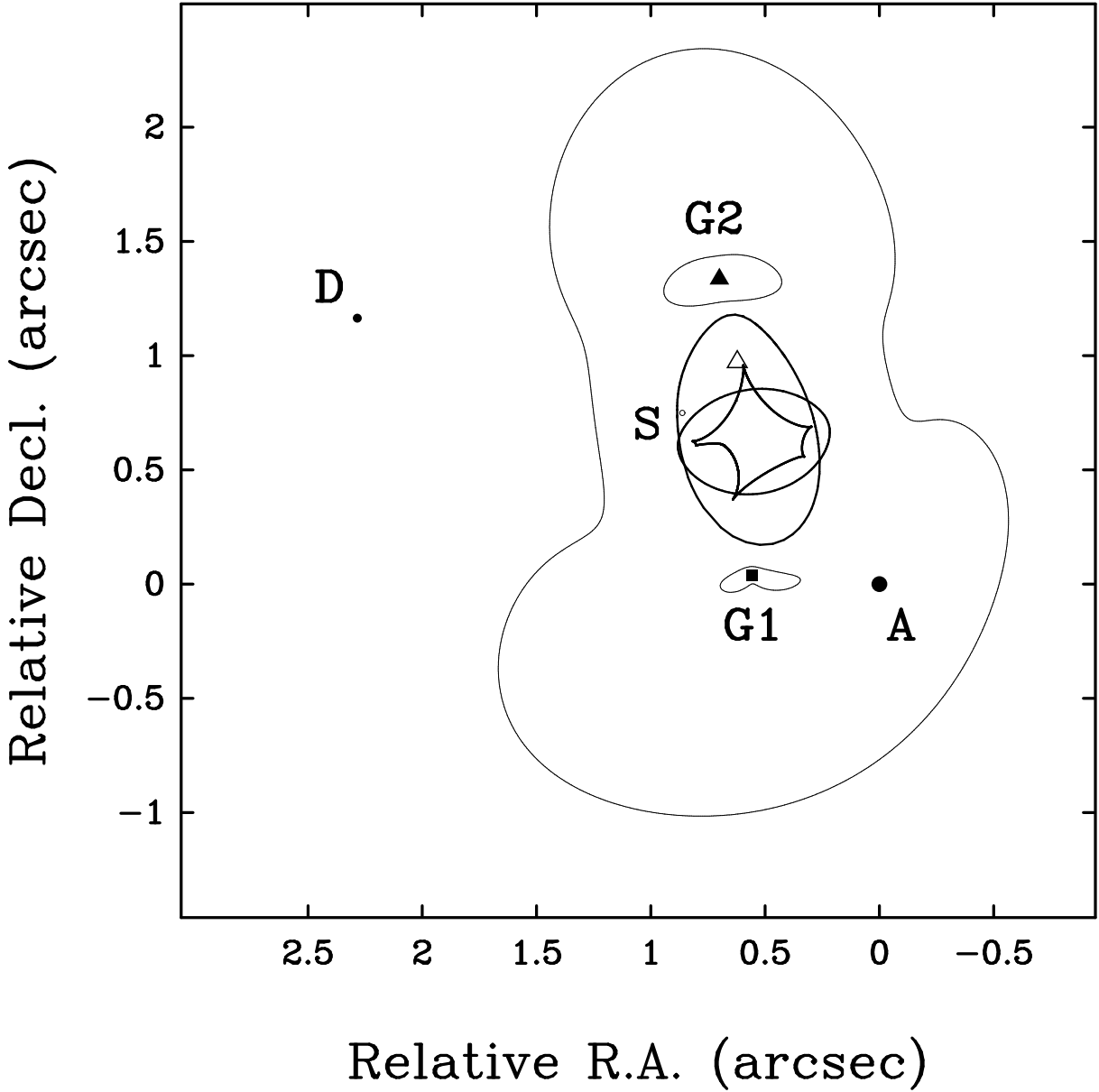}}
\end{picture}
\caption{
An example lens model for radio cores A and D (Model~2; see Table~2, Figure~4b
and Figure~5). The two lensing galaxies are indicated by filled square and 
triangle for G1 and G2, respectively. The `true' position of the background
galaxy is indicated by an open triangle predicted by the model. The predicted
source position of the images is indicated by a small open
circle marked by `S'. The caustics and critical curves for the model
are drawn with thick and thin solid lines, respectively. 
}
\label{}
\end{center}
\end{figure*}

\section{Discussion and Future Work}

In this paper, we have tested and analysed in detail the simplest and
observationally most consistent lensing hypothesis for the JVAS quadruple
radio source B2114+022, namely that the two widest separated, observationally
similar radio components A and D of B2114+022 are lensed images of 
a background radio source due to the two galaxies at different redshifts. 
Although we have tested other lensing scenarios (in which all the four 
components or three components A, B and D are lensed images of a background 
source) and find that they can be consistent with the geometry and radio flux
density ratios, present independent evidence indicating that components B and 
C have different radio surface brightness distributions and radio spectra 
compared to components A and D did not justify detailed analyses of
such scenarios. However, we will have to revisit those lensing 
scenarios in the future if new data and/or new interpretations of the present 
data warrant it. 

Our study finds that the radio components A and D of B2114+022 can be 
successfully reproduced by astrophysically plausible two-plane models 
consistent with the observed properties of the galaxies.
In particular, the unusual flux ratio, i.e. the image closer to the
lensing galaxies being brighter, can be easily reproduced in our model. 
Short radio jets in the north of component A core and those in the south of 
component D core (see Figure~2 of A00) are also consistent with our lens model
prediction; qualitatively, theoretical jets in the south of the radio core on
the source plane can reproduce the jets, based upon our lens models.
Global VLBI observations of these jets in the near future will reveal more
detail and provide more stringent constraints on lens models for B2114+022.
Since the optical counterparts of the radio sources appear to be extremely
faint and are close to the two lensing galaxies (see section~3.1.1), it will
be an observational challenge to obtain optical spectra of the four components
to spectroscopically verify the lensing hypothesis. However, our realistic
model grid (Figure~4) predicts a range of time delays from $\approx 50 - 90$ 
$h_{60}^{-1}$ days between the two components with component D leading 
component A. So if a future radio monitoring program could identify a 
correlated variability between these two components with a time delay, 
this would confirm the lensing hypothesis independently. 
A00 have found no significant variability for the radio sources to date.

Although there are at present only a small number of direct and indirect 
lensing constraints available for B2114+022, the unique geometric arrangement
of the system and the D/A flux ratio appear to allow us to probe galaxies 
mass profiles in terms of radial power-law slope and core radius.
Our study indicates that models adopting two isothermal mass profiles for
the foreground (G1) and background (G2) galaxies potentially may not be a 
particularly good fit to some observed properties of B2114+022. They require 
a relatively high mass ellipticity ($\epsilon \sim 0.5$) for G2 
while the inferred intrinsic light distribution is nearly round
($\epsilon \sim 0.1$). They also imply a much larger
(at least by a factor of $\sim 4$) $M/L$ ratio for G1 than for G2, which
could pose a potential problem for the model.\footnote{The spectra of the
two galaxies show that G1 is a post-starburst elliptical, which A00 classify
as ``E + A'' type, while G2 is a normal elliptical. Potential extra light and
dust in G1 complicate using mass ratios of the galaxies to test lens models.}
These potentially problematic features of the two isothermal galaxies model, 
as we have shown in section~3.2.2, can be avoided if a 
shallower-than-isothermal profile is adopted for G2. However, definitive
bounds could not be put on G2's radial slope 
from lensing analyses of this system mainly because image
splitting occurs with respect to G1's centre than G2's centre
[although the undeflected source position can be closer to G2's centre than
G1's centre (see Figure~6)] and as a result G2's mass profile is less sensitive
to lensing properties. Likewise, no bound on the core radius of G2 can be put
using a theoretical image closest to G2 which may or may not form. Instead,
whether bright theoretical images near G2 can form or not is controlled by
a mass ratio between the two galaxies as illustrated in Figure~4.
Specifically, beyond the upper bound on $f_2^M$, the smaller ellipse-like
caustic in Figure~6 grows to enclose the source, allowing two additional 
images near G2 to form. 

Unlike G2's mass profile, however, definitive limits 
on G1's radial power-law slope ($\nu^{(1)}$) and $\nu^{(1)}$-dependent 
core radius can be put using a theoretical image forming near G1's centre. 
As illustrated in Figure~4b, for a relatively shallow radial profile of G1
a region of parameter space becomes excluded due to a predicted bright image
near G1 whose brightness is insensitive to the core radius. 
As G1's profile gets shallower, this excluded region grows and a profile
shallower than $\nu^{(1)} \approx 1.90$ is virtually excluded for G1 since 
such a profile does not allow a realistic P.A. for G1. In addition to
this definitive lower limit on G1's radial slope, a less strong but likely
upper limit on G1's radial slope is suggested from our study, namely that
a profile  significantly steeper than isothermal is not very likely since
those profiles require mass ellipticities much higher than observed
light ellipticities. Thus, a most likely radial power-law slope for G1 
is a relatively shallow profile just over the lower limit. It is remarkable
that relatively strong bounds can be put on a radial power-law slope of G1 
with only a few number of lensing constraints available for B2114+022. For
models outside the excluded parameter space regions (Figure~4), upper limits
can be put on the core radius of G1 (see Table~2). With more strong lensing
constraints provided by Global VLBI observations of the radio jets of 
B2114+022 A, D in the near future, this system offers a good possibility of 
determining the mass profile of G1 either using a simple model such as 
equation (12) or a more realistic model such as a two-component model 
simulating G1's luminous and dark mass components. We plan to address this in
the future. 

One striking natural consequence of a two-plane lens system (or, any system 
consisting of two close galaxies with different redshifts) that we draw 
attention to in this paper is 
that the background galaxy's light distribution is substantially modified
by the foreground galaxy. The lensing effects are intermediate between weak
lensing and strong lensing and include apparent shift of the galaxy position,
changes of position angle and ellipticity, and distortions of isophotes 
generating arc-like isophotes for outer parts of the optical galaxy. 
While modifications of position angle and ellipticity are not observationally
identifiable lensed features (since intrinsic position angle and 
ellipticity are not measurable), distorted arc-like isophotes
are potentially observationally identifiable lensed features
since they are unique to lensing. Predictions on arc-like isophotes depend on
the mass ratio between the two galaxies (Figure~5).
It will be important to obtain deeper HST images to test
these predictions. Multi-colour images may be particularly useful since
the foreground and background galaxies have different colours, and therefore
it may be easier to identify systematic features in a colour map. Once the
predicted lensed features could be measured in the future, these would provide 
strong constraints on the potential of the foreground galaxy.

Although only one two-plane lens candidate has been discovered so far, we
expect many such cases to be discovered in the future. In particular, 
NGST and SKA will play crucial roles in identifying many two-plane lens 
systems in the optical (Barkana \& Loeb 2000) and radio respectively. 
The prediction that 1--10 per cent of lenses may be
two-plane lenses by Kochanek \& Apostolakis (1988) is based on simple
spherical lenses, it would be important to revisit the problem
using more realistic lens models; we plan to address this question in a
further work, with particular emphasis on how this fraction depends on
cosmologies. 

This study shows that the caustics and critical curves of two
elliptical deflectors at different redshifts are extremely complicated
(see Figure~6); these caustics include regions that can produce
7 or 9 images (with the central image strongly demagnified), similar 
to binary galaxies in single-plane lensing (Keeton, Mao \& Witt 2000). 
The effects of additional images and moderate changes in isophotal shapes
can potentially provide strong constraints on lens model parameters and
perhaps cosmological parameters as well. 
Some of these effects were illustrated in this paper. In particular,
the effects of additional images were used to put model-dependent limits on 
the mass ratio of the two galaxies (see Figure~4) and to limit the allowed
range of the radial power-law slope of the foreground galaxy. This can be 
understood as follows. In the complicated caustic structure of
the two plane lens (see Figure~6), the caustics close to the source
are more sensitive to the change of some model parameters
than in a simpler caustic structure lens. An idealized example further
illustrates the potential power of two plane lenses as astrophysical tools. 
If we have two perfectly aligned galaxies at different redshifts lensing
a distant source, the aligned background source will be imaged
into two Einstein rings while the background galaxy is imaged into a
third ring. If we model the lensing galaxies as singular isothermal spheres,
then there are only two velocity dispersions that parameterize the lenses,
while we have three Einstein ring size constraints. The one extra
constraint can then be used to constrain other parameters such as cosmology.
It will be very interesting to see in the near future whether two-plane lenses
can be a robust tool for cosmological studies.

\section*{Acknowledgments}
We thank Neal Jackson for providing us with Figure~3. We are grateful to
him, Ian Browne, Chuck Keeton and Peter Wilkinson for encouragements and
many helpful discussions.

{}

\bsp
\label{lastpage}

\begin{thebibliography}{}
\bibitem{} Arnaboldi M., Cappaccioli M., Cappellaro E., Held E. V., Sparke L.,
  1993, A\&{A}, 267, 21
\bibitem{} Augusto P., et al., 2000, MNRAS, submitted (A00)
\bibitem{} Barkana R., Loeb A., 2000, ApJ, 531, 613
\bibitem{} Blandford R., Narayan R., 1986, ApJ, 310, 568
\bibitem{} Browne I. W. A., 2000, in Gravitational Lensing: Recent
        Progress and Future Goals, ASP conference series, eds. T. G. 
        Brainerd \& C. S. Kochanek, in press
\bibitem{} Buote D. A., Canizares C. R., 1997, in Galactic Halos: A UC Santa 
 Cruz Workshop, ASP Conference series Vol.\ 136, ed.\ D. Zaritsky
\bibitem{} Chae K.-H., Khersonsky V. K., Turnshek D. A., 1998, ApJ, 506, 80
\bibitem{} Erdl H., Schneider P., 1993, A\&A, 268, 453
\bibitem{} Keeton C. R., Kochanek C. S., Falco E. E., 1998, ApJ, 509, 561
\bibitem{} Keeton C. R., Kochanek C. S., Seljak U., 1997, ApJ, 482, 604
\bibitem{} Keeton C. R., Mao S., Witt H. J., 2000, ApJ, 537, 697
\bibitem{} Kochanek C.S., Apostolakis J., 1988, MNRAS, 235, 1073
\bibitem{} Kovner I., 1987, ApJ, 316, 52
\bibitem{} Norbury M., et al., 2000, MNRAS, submitted
\bibitem{} Sackett P. D., 1999, in Galaxy Dynamics, 
 ASP Conference series Vol.\ 182, eds.\ D. R. Merritt, M. Valluri and J. A.
 Sellwood
\bibitem{} Sackett P. D., Rix H.-W., Jarvis B. J., Freeman K. C., 1994, ApJ,
    436, 629
\bibitem{} Sackett P. D., Pogge R. W., 1995, in Dark Matter, AIP Conference 
        series Vol.\ 336, eds.\ S. S. Holt \& C. L. Bennett
\bibitem{} Schneider P., Ehlers J., Falco E.E., 1992, 
  Gravitational Lenses (New York: Springer-Verlag)
\bibitem{} Seitz S., Schneider P., 1992, A\&A, 265, 1
\end{thebibliography}
\end{document}